# Thermodynamic properties of finite binary strings

Sergei Viznyuk

**Abstract**

Thermodynamic properties such as temperature, pressure, and internal energy have been defined for finite binary strings from equilibrium distribution of a chosen computable measure. It is demonstrated a binary string can be associated with one-dimensional gas of quasi-particles of certain mass, momentum, and energy

A variety of entities such as complex computer and social networks, World Wide Web, ecological systems, other *scale-free networks* [1], have been found to exhibit behavior closely described by laws of statistical mechanics and thermodynamics [2-5]. *Ising models* applied to such systems predicted phenomena resembling Bose-Einstein condensation [5] and ferromagnetic phase transition [6]. Binary strings are another type of objects which are common target of statistical analysis [7-8]. The equilibrium temperature, and partition function for pairs of binary strings have been previously formulated [8] based on definition of interaction energy as *Hamming distance* [9] between different strings in the ensemble.

We demonstrate thermodynamic properties can be calculated for any binary string of finite length. The equilibrium distribution defines the energy levels and equilibrium temperature.

Previously it was suggested [10] to associate an ensemble of strings **C** with a given binary string **B** through a computable measure $C_n$ which is the *Hamming distance* between **B** and string **B** itself shifted by *n* bits:

$$C_n = \sum_{i=1}^{M} B_i \oplus B_{i+n} \ , \ \ 0 \leq n < N \ , \ \ 0 < N \leq M \tag{1}$$

, where $B_i$ is $i^{th}$ bit of a binary string **B**, consisting of *M* bits with values 0 or 1, and $\oplus$ is logical XOR operator; $B_{i+n} = B_{i+n-M}$ for $i+n>M$; *N* is the total number of strings in ensemble **C**. The product $B_i \oplus B_{i+n}, 1 \leq i \leq M$ is a string $C_n$ of *M* bits. The value $C_n$ is the sum of set bits in string $C_n$. We take $C_n$ values as *the observable measure* of the ensemble **C**, and *N* as the total number of observations. Following are the associated measures and properties:

1) $0 \leq C_n \leq 2 \times \min(k, M-k)$, where *k* is the number of set bits in string **B**

2) $0 < N_i < M$, where $N_i$ is the number of occurrences of $i^{th}$ distinct value $C_i$ in the ensemble; $0 < i \leq 1 + \min(k, M-k)$

3) $\sum_i N_i = N$

4) $\overline{C} = \frac{1}{N}\sum_i N_i \cdot C_i = \frac{1}{M}\sum_{n=0}^{M-1} C_n = 2 \cdot M \cdot p \cdot (1-p)$, where $\overline{C}$ is the average value of $C_n$ in the ensemble; $p = \frac{k}{M}$; *k* is the number of set bits in string **B**.

From *central limit theorem* we expect the distribution $N_i(C_i)$ for random strings to converge to normal distribution as length $M$ increases:

$$N_i^{random} \cong N_i^{normal} = \frac{2 \times N}{\sqrt{2 \cdot \pi \cdot \sigma^2}} \exp\left(-\frac{(C_i - \overline{C})^2}{2 \cdot \sigma^2}\right) = N_0 \exp\left(-\frac{(C_i - \overline{C})^2}{2 \cdot M \cdot K \cdot \frac{\overline{C}}{M} \cdot \left(1 - \frac{\overline{C}}{M}\right)}\right) \quad (2)$$

, where $N_0 = \frac{2 \times N}{\sqrt{2 \cdot \pi \cdot \sigma^2}}$ , variance $\sigma^2 = M \cdot K \cdot \frac{\overline{C}}{M} \cdot \left(1 - \frac{\overline{C}}{M}\right)$ , $K = 1 - \sqrt{\left|1 - 2 \cdot \frac{\overline{C}}{M}\right|}$

Formula (2) is an approximation of adjusted binomial distribution

$$N_i^{binomial} = \frac{2 \times N}{K} \cdot \left(\frac{\left(\frac{M}{K}\right)!}{\left(\frac{C_i}{K}\right)! \left(\frac{M - C_i}{K}\right)!} \cdot \left(\frac{\overline{C}}{M}\right)^{\frac{C_i}{K}} \cdot \left(1 - \frac{\overline{C}}{M}\right)^{\frac{M - C_i}{K}}\right) \quad (3)$$

Note, that factor 2 in front of $2 \times N$ above and in property 4) appears because the same $C_i$ value can be realized with $k$ and $M$-$k$ set bits in string **B**.

Figure 1 shows distributions $N_i(C_i)$ calculated using (1) for four different strings **B**:

1. English text of Einstein's address to Prussian Academy of Sciences, *Principles of Theoretical Physics*, 1914, 6.6kB in length, blue color dots.

2. The same text compressed with 'gzip -9', 3kB, green color dots.

3. UNIX (AIX 5.3 64-bit) executable file, 38kB, red color dots.

4. Random string produced as output from /dev/urandom on UNIX, 16kB, purple color dots.

The $N_i(C_i)$ distributions are shown as dots. The corresponding normal distributions calculated using (2) are plotted as lines.

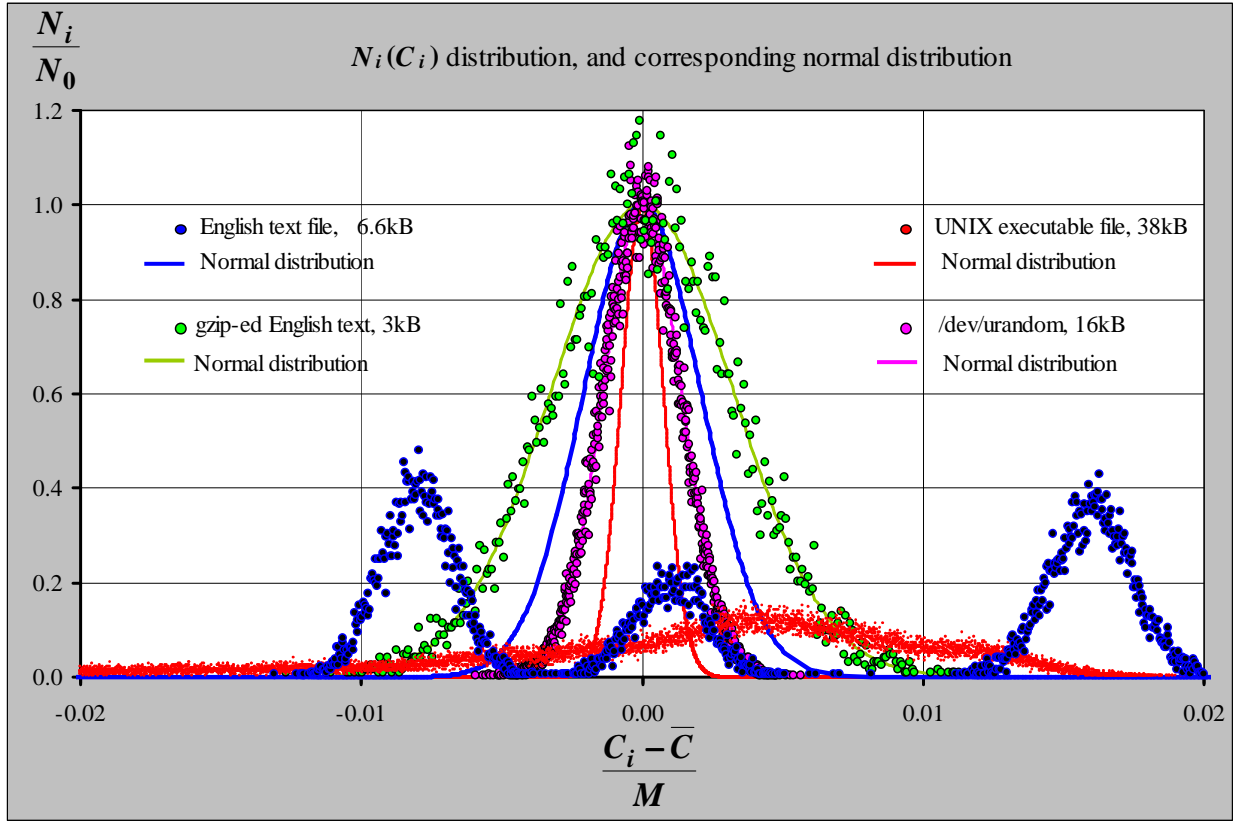

Figure 1

The normal distribution on the graph is virtually identical to the adjusted binomial (3) for all four cases. The graphs demonstrate $N_i(C_i)$ distribution for random string and compressed string is well approximated by normal distribution. We consider such distribution as *equilibrium state* of ensemble $C$. The $N_i(C_i)$ distribution for non-random strings, such as English text and UNIX executable file differ significantly from corresponding normal distribution. We consider such ensemble $C$ to be in *non-equilibrium state*. As follows from (2) the equilibrium distribution is provided by

$$N_i^{eq} = N_i^{random} \cong N \cdot \sqrt{\frac{2}{\pi \cdot M \cdot T}} \cdot \exp\left(-\frac{E_i}{T}\right) \qquad (4)$$

The equilibrium distribution (4) gives the definition of energy levels

$$E_i = \frac{(C_i - \overline{C})^2}{2 \cdot M} = \frac{P_i^2}{2 \cdot M} \qquad (5)$$

, where $P_i = C_i - \overline{C}$ can be considered particle momentum, and $M$ as the mass of the particle. The "particles" here are the strings $C_n$ in ensemble defined by (1). The equilibrium temperature is thus provided by

$$T = K \cdot \frac{\overline{C}}{M} \cdot \left(1 - \frac{\overline{C}}{M}\right) = \left(1 - \sqrt{\left|1 - 2 \cdot \frac{\overline{C}}{M}\right|}\right) \cdot \frac{\overline{C}}{M} \cdot \left(1 - \frac{\overline{C}}{M}\right) \qquad (6)$$

. The average per particle internal energy can then be calculated as

$$\overline{U} = \frac{1}{N}\sum_i N_i \cdot E_i = \frac{1}{N} \cdot \sum_i N_i \cdot \frac{(C_i - \overline{C})^2}{2 \cdot M} \qquad (7)$$

, and average in bits per particle entropy as

$$\overline{S} = \frac{1}{N}\log_2(\Omega) = \frac{1}{N}\log_2\left(\frac{N!}{\prod_i N_i!} \times \prod_n \left(2 \cdot \frac{M!}{C_n!(M-C_n)!}\right)\right) =$$

$$= \frac{1}{N} \cdot \left(\log_2(N!) - \sum_i \log_2(N_i!)\right) + \frac{1}{N}\sum_i N_i \cdot \log_2\left(2 \cdot \frac{M!}{C_i!(M-C_i)!}\right) =$$

$$= 1 + \frac{\log_2(e)}{N} \cdot \left(\log_e(\Gamma(N+1)) - \sum_i \log_e(\Gamma(N_i+1))\right) + \qquad (8)$$

$$+ \frac{\log_2(e)}{N}\sum_i N_i \cdot \left(\log_e(\Gamma(M+1)) - \log_e(\Gamma(C_i+1)) - \log_e(\Gamma(M-C_i+1))\right) =$$

$$= \overline{S}_{thermodynamic} + \overline{S}_{microstate}$$

, where

$$\overline{S}_{thermodynamic} = \frac{\log_2(e)}{N} \cdot \left(\log_e(\Gamma(N+1)) - \sum_i \log_e(\Gamma(N_i+1))\right)$$

$$\overline{S}_{microstate} = 1 + \frac{\log_2(e)}{N}\sum_i N_i \cdot \left(\log_e(\Gamma(M+1)) - \log_e(\Gamma(C_i+1)) - \log_e(\Gamma(M-C_i+1))\right)$$

Modern programming languages have $\log_e(\Gamma(x))$ function defined, for example *lgamma(x)* in C. Therefore the calculation of entropy from (8) and internal energy from (7) is straightforward given known distribution $N_i(C_i)$ obtained from (1), for arbitrary binary string **B** of finite length. Calculation of thermodynamic quantities for equilibrium state can be done analytically, based on partition function, known from (4)

$$Z_N = \sum_i \exp\left(\frac{E_i}{T}\right) = \sqrt{\frac{\pi M T}{2}} \qquad (9)$$

The average equilibrium internal energy per particle is calculated using (9) as

$$\overline{U}^{eq} = T^2 \frac{\partial}{\partial T}(\log_e Z_N) = T^2 \frac{\partial}{\partial T}\log_e\left(\sqrt{\frac{\pi M T}{2}}\right) = \frac{T}{2} \qquad (10)$$

This result is consistent with average energy per particle in one-dimensional Maxwell-Boltzmann gas. The average equilibrium entropy in bits per particle is

$$\overline{S}^{eq}_{ther\,mod\,ynamic} = \log_2(e) \cdot \frac{\partial}{\partial T}(T \cdot \log_e(Z_N)) = \frac{1}{2} \cdot \log_2\left(\frac{\pi \cdot e \cdot M \cdot T}{2}\right)$$

$$\overline{S}^{eq}_{microstate} \cong \overline{C} \cdot \log_2\left(\frac{M}{\overline{C}}\right) + (M - \overline{C}) \cdot \log_2\left(\frac{M}{M - \overline{C}}\right)$$

(11)

Other thermodynamic relations follow; *S*, *U,* and *F* are the entropy in *nats*, internal energy, and Helmholtz free energy for the whole ensemble:

$$S = \frac{N}{2} \cdot \log_e\left(\frac{\pi \cdot e \cdot V^2 \cdot T}{2}\right)$$

$$F = U - TS = -\frac{N \cdot T}{2} \log_e\left(\frac{\pi \cdot V^2 \cdot T}{2}\right)$$

$$P = -\left(\frac{\partial F}{\partial V}\right)_{T,N} = \frac{N \cdot T}{V}$$

(12)

, where "volume" $V = \sqrt{M}$ ; *P* is the pressure. For ensemble *C* built using (1) from random source string *B* which has the probability of set bit $p = \frac{1}{2}$ :

$$T = K \cdot \frac{\overline{C}}{M} \cdot \left(1 - \frac{\overline{C}}{M}\right) = \frac{1}{4}$$

$$\overline{U}^{eq} = \frac{1}{8}$$

(13)

Table 1 presents results of computation of average internal energy and entropy in bits per particle for four ensembles *C* built from four strings *B* used for graphs on Figure 1. Computations were performed using (7) and (8). The corresponding equilibrium average internal energy and entropy have been calculated using (10) and (11). The microstate entropy is in bits per bit.

| ***B*** string sample | $\overline{U}$ | $\overline{U}^{eq}$ | $\overline{S}_{thermo}$ | $\overline{S}^{eq}_{thermo}$ | $\overline{S}_{micro}/M$ | $\overline{S}^{eq}_{micro}/M$ |
|---|---|---|---|---|---|---|
| English text, 6.6kB | 131.61 | 0.1234 | 9.447 | 7.833 | 0.9853 | 0.9999 |
| Compressed English text, 3kB | 0.1265 | 0.1250 | 7.259 | 7.308 | 0.9997 | 1.0000 |
| UNIX executable file, 38kB | 20.455 | 0.1044 | 12.335 | 8.921 | 0.9942 | 0.9946 |
| Random string, 16kB | 0.1246 | 0.1250 | 8.498 | 8.528 | 0.9999 | 1.0000 |

Table 1

The results of calculations show match of values $\overline{U}^{eq}$ and $\overline{S}^{eq}_{micro}$ for compressed file and random file with (13) to within 0.01%.

The presented model can be applied to a pairs of different source strings (***A***, ***B***), not necessarily of the same length, using the following ensemble definition:

$$C_{(A,B)} = \sum_{i=1}^{M} C_i = \sum_{i=1}^{M} A_m \oplus B_n$$

(14)

, where $m=(i-1)\% M_A$ , $n=(i-1)\% M_B$ , $M = lcm(M_A, M_B)$.

Here $M_A$ and $M_B$ are the lengths of strings *A* and *B* in bits, $i\%M$ is remainder of division of *i* by *M*, lcm($M_A, M_B$) is the *least common multiple* of $M_A$ and $M_B$.